\documentclass[sigconf]{acmart}

\usepackage{balance} 
\usepackage{listings}
\usepackage{subcaption}

\title[Cohorts and Dense Urban Metro Networks]{Cohorting to isolate asymptomatic spreaders: An agent-based simulation study on the Mumbai Suburban Railway}

\author{Alok Talekar$^1$, Sharad Shriram$^2$, Nidhin Vaidhiyan$^2$, Gaurav Aggarwal$^1$, Jiangzhuo Chen$^3$, Srini Venkatramanan$^3$, Lijing Wang$^3$, Aniruddha Adiga$^3$, Adam Sadilek$^1$, Ashish Tendulkar$^1$, Madhav Marathe$^3$, Rajesh Sundaresan$^{2,4}$ and Milind Tambe$^1$}
\affiliation{
  \institution{
        $^1$ Google Inc.\\
        $^2$ Indian Institute of Science, Bangalore\\
        $^3$ University of Virginia \\
        $^4$ Strand Life Sciences
    }
  }
  
\begin{abstract}
The Mumbai Suburban Railways, \emph{locals}, are a key transit infrastructure of the city and is crucial for resuming normal economic activity. 
Due to high density during transit, the potential risk of disease transmission is high, and the government has taken a wait and see approach to resume normal operations. 
To reduce disease transmission, policymakers can enforce reduced crowding and mandate wearing of masks.
\emph{Cohorting} -- forming groups of travelers that always travel together, is an additional policy to reduce disease transmission on \textit{locals} without severe restrictions. Cohorting allows us to:
($i$) form traveler bubbles, thereby decreasing the number of distinct interactions over time; ($ii$) potentially quarantine an entire cohort if a single case is detected, making contact tracing more efficient, and ($iii$) target cohorts for testing and early detection of symptomatic as well as asymptomatic cases.
Studying impact of cohorts using compartmental models is challenging because of the ensuing representational complexity. Agent-based models provide a natural way to represent cohorts along with the representation of the cohort members with the larger social network. This paper describes a novel multi-scale agent-based model to study the impact of cohorting strategies on COVID-19 dynamics in Mumbai.
We achieve this by modeling the Mumbai urban region using a detailed agent-based model comprising of 12.4 million agents.
Individual cohorts and their inter-cohort interactions as they travel on locals are modeled using local mean field approximations. The resulting multi-scale model in conjunction with a detailed disease transmission and intervention simulator is used to assess various cohorting strategies. The results provide
a quantitative trade-off between cohort size and its impact on
disease dynamics and well being. The results show that cohorts can provide significant benefit in terms of reduced transmission
without significantly impacting ridership and or economic \& social activity.
\end{abstract}


\renewcommand\footnotetextcopyrightpermission[1]{} 
\begin{document}

\maketitle 
\pagestyle{plain} 
\section{Introduction}

\begin{figure*}[t]
  \centering
  \includegraphics[width=\linewidth]{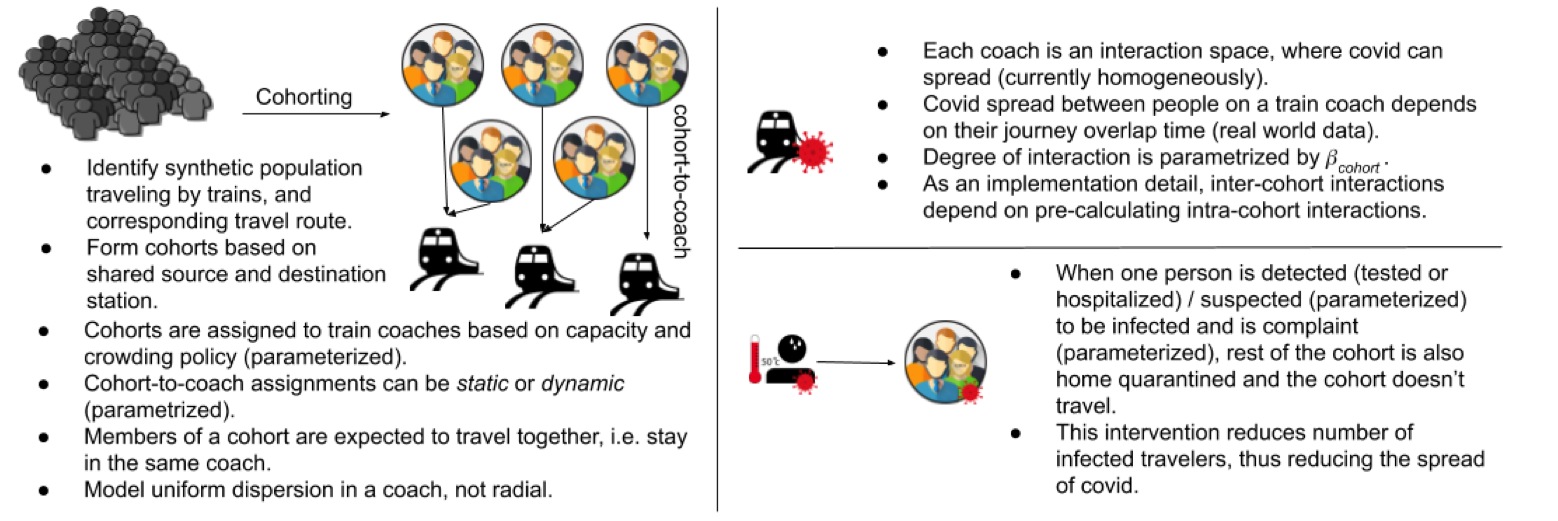} 
  \caption{Schematic representation of our agent-based modeling of local trains.}
  \label{fig:logo}
  \Description{Schematic of agent-based modeling with cohorts.}
\end{figure*}

COVID-19 is the worst public health disaster in the 21st century, second only to the 1918 pandemic in the last 100 years. An estimated $36$ million confirmed cases and over a million fatalities all over the world have been reported as of $8$th October $2020$. The response to the pandemic has varied across countries, but given that no pharmaceutical interventions were available, almost all countries instituted significant social distancing measures to control the spread.  As an extreme measure, countries had enforced lockdowns to reduce mobility in varying capacity. Data clearly shows that urban regions 
were impacted far earlier than rural regions. Population density and the resulting social interactions are important driving factors.

The pandemic has had significant impact on India thus far.
The Government of India acted early and instituted a nation-wide lockdown. The lockdown had a significant effect on
urban and regional mobility patterns\footnote{See \href{https://ourworldindata.org/covid-mobility-trends}{https://ourworldindata.org/covid-mobility-trends.}. Our results estimate that there was over 70\% drop in mobility in India \cite{adiga2020interplay}}. Like other parts of the world, Indian
cities, many of which have some of the highest population densities, have been affected significantly by the pandemic. 
Mumbai, one of the largest metropolises in the world with 12.4 million inhabitants in the Brihanmumbai area\footnote{This is the area under the jurisdiction of the Brihanmumbai Municipal Corporation.} is the economic hub of the country. Mumbai has the highest population density in the country (over $32,300\, people/km^2$), and has been significantly impacted by the pandemic.
\textit{Locals} had a daily ridership of over $8.2$ million passengers before the lockdown -- this represents 20-40\% of city's total population\footnote{The estimate accounts for multiple rides by the same individual}. 
\emph{Locals}, the Mumbai suburban trains, are a lifeline of the city and play a central role in the vibrant social and economic activities of the metropolis, and resumption of service is critical for the resumption of normal economic and social activities in Mumbai. However, this can potentially contribute to a surge in disease spread \cite{harris2020subways} since the locals and the stations are densely packed during normal working hours. This creates a dense social network, ideal for airborne disease transmission, with long distance edges that can quickly spread the disease across the metropolis. The source code for the simulator is open-sourced and available at\\ \href{https://github.com/cni-iisc/epidemic-simulator/tree/mumbai\_local}{\textcolor{blue}{https://github.com/cni-iisc/epidemic-simulator/tree/mumbai\_local}}
\subsection{Our contributions and novelty} 
This is the first paper to explore and quantify the public health benefit of {\em cohorting} strategies while facilitating increased mobility and economic activity. Policy makers are looking for ways to resume the service of {\em locals} in Mumbai and this work enables just that.
Cohorts \footnote{An illustrative examples is shown \href{https://youtu.be/6H8hZBNcP8k}{\textcolor{blue}{on this video}}} are motivated by two central hypotheses. ($i$) Two social interaction networks with the same interaction density can have varying impacts on disease dynamics; specifically, for small social networks,
locally dense but globally weakly connected networks might be better at controlling the spread. ($ii$) It is easy to do contact tracing when travelers form cohorts; early and efficient contact tracing followed by quarantining and monitoring of individuals who came in close contact with the infected individuals can help mitigate the disease spread. 
The first hypothesis is the rationale for \emph{social bubbles} \cite{nielsen2020covid} in the context of restricted social interactions. From this perspective, cohorts can be thought as {\em traveler bubbles}. They change the dynamic social network structure by creating locally dense sub-graphs but reducing interactions between these dense sub-graphs.
The second hypothesis is that cohorting helps us implement the {\em test-isolate} \cite{whoPressRelease} strategy with greater efficiency. Together, cohorting can potentially help reduce the overall impact of the epidemic while supporting social and economic activities --- the precise trade-offs of these two components is the focus of the paper.\\

\noindent
\textbf{Novelty.} \
The work has a number of innovative components. This includes the following.
($i$) \emph{Multi-scale agent simulations:} \ representing cohorts within
a social network is challenging and calls for a multi-scale approach -- agents interacting through interaction spaces instead of interacting directly with each other -- in order to reduce computational complexity without compromising the quality of the solution. 
Another multi-scale notion is that of hierarchical interaction spaces that stratify interactions within a larger interaction space like in a workplace or a school or a neighbourhood. This is essential for simulating contact tracing and the test-isolate strategy effectively.
($ii$) \emph{Multi-theory models:} \ that incorporate multiple social and epidemic theories. 
This includes theories of disease transmission and social choice theories (route choices to workplaces).
($iii$) \emph{Calibration to diverse data sets.} \ Agent models should be calibrated to complex measured data. We incorporate a large variety of data to calibrate and validate the models. 
($iv$) \emph{Evaluation of realistic policies:} \ 
Study of policies that are directly derived from needs on the ground. 
The public health outcomes we derive from such a data driven model provides innovative insights that can inform policy.

\section{Related Work}
The literature on agent-based models (ABMs) is vast, see for example \cite{del2006episims,barrett2008episimdemics,balmer2008agent} (ABM frameworks), \cite{eubank2004modelling} (modeling of small pox spread), \cite{hackl2019epidemic} (benefit of fine-grained model), \cite{chen2016effect,adiga2018disparities} (fine-grained modeling of slums in Delhi and influenza spread), \cite{venkatramanan2018using} (benefit of using auxiliary data sources for modeling Ebola spread), \cite{singh2018behavior}  (use of survey data to model behaviour),  \cite{antelmi2020design, tizzani2018epidemic, vestergaard2015temporal} (time-varying graph networks), \cite{sambaturu2020designing} (vaccination), \cite{swarup2014computational} (ABMs for computational epidemiology), \cite{pastor2015epidemic, marathe2013computational} (review of epidemic spread in complex networks) and \cite{wang2020nyu} (impact of mobility on COVID-19 spread). 

We shall focus only on those closely related to our work in this paper. Mei et al. \cite{mei2015simulating} developed a city-scale ABM for Beijing (6M agents) using detailed demographic, mobility, socio-economic data. The data they needed are quite detailed, the interventions modeled have limited flexibility, and the test-isolate strategy is not modeled. Cooley et al. \cite{cooley2011role} developed a city-scale ABM (7.5M  agents) based on survey data to study H1N1 epidemic spread in the New York City. They calibrated the parameters based on the 1957-8 influenza spread data. The interventions are limited to wearing masks, vaccinations and social distancing. They conclude that interventions targeting the subway system alone are not sufficiently effective in mitigating the infection spread. The papers \cite{harris2020subways, glaeser2020princeton, gosce2018londonTube} use a density model developed in \cite{gosce2014corridor} for narrow and enclosed areas to study the correlation between mobility and infection spread in crowded spaces. The validation of these models are limited to statistical data analysis \cite{harris2020subways} and small-scale simulations (200 -- 100,000 agents) \cite{gosce2014corridor, gosce2018londonTube}. Social bubbles have been studied by \cite{nielsen2020covid, pinto2020local,leng2020effectiveness} using ABMs. However, these ABMs are modeled with limited interaction spaces \cite{leng2020effectiveness}, or do not consider local demographic variations \cite{pinto2020local}, or limit the number of agents to about 100,000 \cite{block2020social}. 

In this work, we overcome several of the limitations highlighted above. We model the interactions more realistically by considering multiple and hierarchical interactions spaces like homes/neighbourhoods/communities, schools, and workplaces, in addition to interactions during commute. Our simulator works with 12.4M agents. The contact rate parameters are calibrated to the COVID-19 India data. The simulator enables time-varying interventions that model on-ground policies (containment zones, lockdown fatigue, compliance,  contact tracing policies, and quarantine duration). Additionally, the simulator implements the test-isolate strategy.
\section{Methodology}
We implement cohorting strategies on top of a city-scale agent-based epidemic simulator. The agent-based simulator models interactions in households, workplaces, schools, neighbourhoods, and communities. In this paper, we focus on the modeling and the implementation of cohorts as an additional interaction space related to transportation. 

The agent-based simulator consists of two components: ($i$) a synthetic city generator and ($ii$) a disease spread simulator that simulates the spread of the infection on the generated synthetic city. The synthetic city is generated taking into account the demographics of the city such as: age distribution, household size distribution, unemployment ratio, commute distance distributions, school size distributions etc. The synthetic city has as many agents as the population in the city of interest.


An attribute particularly important to the current study is the workplace location of an individual. We first assign a workplace ward (Mumbai has 24 wards) to an individual in accordance with the inter zone travel pattern given in Table 4 in \cite{ref:mumbai_commute_world_bank_report}. The individual is then randomly assigned a workplace in the workplace ward (workplace location is uniformly distributed across a ward), with the intention to match the commute distance distributions for Mumbai.

All interactions between agents are modeled via  interaction spaces to enable scalable computation at city-scale. Examples of interaction spaces include train coaches, workplaces, households, neighbourhoods. An agent can have membership to multiple interaction spaces, and disease spreads between agents via shared interaction spaces. Additionally, we bring hierarchical interactions spaces (project-team : workplace, neighborhood : wider-community, class-rooms : schools) to enable modeling of test-isolate strategies. The infection spreading model for an interaction space can depend on the type of the interaction space.

The model for disease progression in an infected individual accounts for disease states such as susceptible, exposed, pre-symptomatic and asymptomatic (to model asymptomatic transmission), symptomatic, hospitalised, critical, deceased and recovered, with age-dependent state transition probabilities. The age-dependent state transition probabilities are based on estimates in \cite{verity2020estimates}.

With all these attributes, agents constitute the nodes of a geo-spatial social contact network with mapped mobility. The city-scale epidemic simulator takes $(i)$ the instantiated network (synthetic city), $(ii)$ the disease progression model, $(iii)$ the intervention, testing and contact tracing protocols, and then simulates the spread of infection in the network. We expand on the modeling of the transport interaction space and cohorts in the next subsections.

\subsection{Mumbai \emph{locals} Dataset}
We digitized the Mumbai Rail Map in \cite{ref:mumbai_rail_map}, and captured train line, train station, transit time between consecutive stations along a line, along with station latitude, longitude information. We restrict ourselves to the census city limits of Mumbai and Mumbai suburbs, to be consistent with agents being generated, and have 52 stations in the network. This data is used to pre-compute shortest travel route and corresponding time between any two stations across any line. In determining total travel time, if a journey has multiple legs, each transfer interval is assumed to be 7 minutes based on frequency of \emph{locals} at stations under pre-COVID schedule. A leg represents transit along a single train line that does not require any transfer.
During instantiation, the pre-computed travel times are used in determining shortest commute time route and corresponding source-destination stations for individuals. In the simulator, the pre-computed travel times are used to model the time spent in a journey.

\subsection{Cohorts}
\label{ref:Cohorts}
To model cohorts, in addition to the attributes mentioned above, we first determine if an office-going agent would take the train to go to work. For travel between home and workplace we optimize for travel time across various modes of travel, accounting for frequency and cost differential of road commute options compared to using \emph{locals}, the expected speed of travel via road 21.6 km/h \cite{ref:mumbai_avg_speed} along with \href{https://www.ncbi.nlm.nih.gov/pmc/articles/PMC3835347/}{geodesic to road detour index} \cite{distanceaFactor} (value considered for Mumbai = 1.7), and multiple possible train stations an agent would consider as the primary commute stations both for home and workplace location. This allows us to know which agents take trains, and what route they follow. Our generated data shows that about 30\% of the population takes train for daily workplace commute, which is inline with the known  ridership numbers of \emph{locals}. Each cohort can have up to 3 legs in the journey (based on \emph{locals} routes), where each leg represents travel along a single line of \emph{locals} in a single coach.

All individuals in a cohort are assumed to travel together while commuting. Hence they are assigned based on shared origin and destination stations, that is every member of the cohort would assemble at the common origin station, and then travel together to the destination station together with other members of the cohort and remain in the same train coach for each leg of the journey as every other member of the cohort. In our simulation, individuals are picked randomly to form a cohort as long as they share origin and destination stations for commute. Cohort size is parameterized, and the case of cohort size = 1 represents non-cohorting (business as usual) scenario. Cohorts are formed at the start of the simulation and stay same throughout the simulation. 

\subsubsection{Cohort-to-coach matching}
Each day cohorts get assigned to train coaches for their morning and evening commute. For every train line (four in our case) and direction of travel, we initialise an empty coach with a given seating capacity and a \emph{crowding factor}. For any given section of travel, the number of individuals in a coach cannot exceed the occupancy limit defined as the product of seating capacity and the crowding factor. In other words, the crowding factor captures the fact that the number of travelers in a single coach almost exceeds the total number of available seats (by a significant factor during normal operations). Our experiments
use crowding factor as one of the parameters that can be controlled when formulating a policy. For each coach, we also keep count of cohorts who could not be accommodated due to unavailability of space. Next, we pick cohorts at random. For the chosen cohort, we check if the existing coaches on the journey legs for the cohort (depending on their source destination pair, cohorts can travel across multiple lines and hence could have more than one journey leg) can accommodate the cohort. If the cohort can be accommodated on all its journey legs, the cohort is assigned to the corresponding coaches, and the capacity of the coaches along the cohort's journey legs is correspondingly reduced. If the cohort cannot be accommodated in at least one journey leg, the cohort is put back into bin of un-allocated cohorts, and the counter on coaches that could not accommodate the cohort is incremented. If the counter on a coach exceeds a threshold (five in our simulations), the current coach is pushed into an array of occupied coaches, and a new coach is instantiated for that train line. We continue this till all cohorts are assigned to coaches.\\

\begin{figure}[t]
  \centering
  \includegraphics[width=0.7\linewidth]{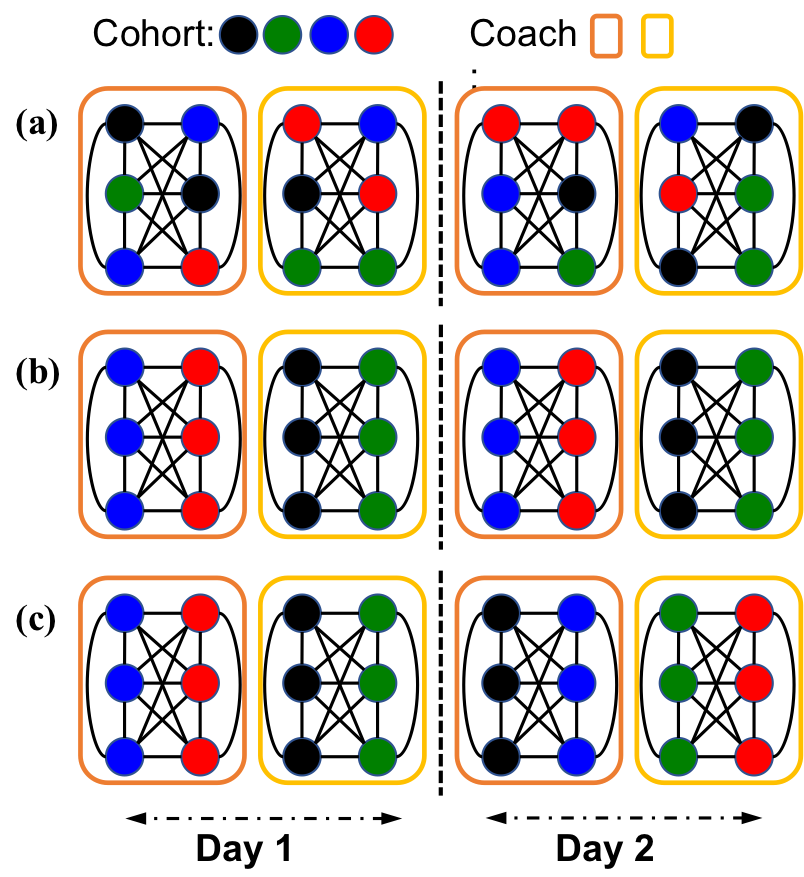} 
  \caption{An example of different cohort-to-coach strategies and their impact on network structure. There are four cohorts (colored by black, green, blue, and red) and two coaches (colored by orange and yellow). Two days are shown in the figure. (a) No cohort-to-coach strategy. Passengers from all cohorts are randomly mixing between coaches on different days. (b) Static cohort-to-coach strategy. Members of a cohort travel together and coach assigned to the cohort is fixed across days. (c) Dynamic cohort-to-coach strategy. Members of a cohort travel together and coach assigned to the cohort is arbitrary across days. }
  \label{fig:cohort}
\end{figure}

{\bf Assigning cohort-to-coach.} \ 
We study two cohort-to-coach assignment strategies -- static and dynamic.
Static assignment strategy is used to model the case where the coach assigned to a cohort is fixed for each commute via a ticketing policy in practice. 
Specifically, each cohort takes the \emph{same coach each day} in each direction, and this coach is shared with the \emph{same cohorts} each day as well. 
Dynamic assignment strategy is used to model the case where the ticketing policy is more relaxed, and each cohort can potentially choose any train coach arbitrarily as long as they meet the basic criteria of cohorting where every member of the cohort travels together in the same coach. The dynamic strategy is easier to implement and requires minimal enforcement in practice compared to static assignment. The impact of the strategies
is evaluated and discussed in the paper later. Figure~\ref{fig:cohort} illustrates the cohorting strategies and their impact on network structure.

\subsubsection{One-Off Travel}
We model one-off travel for scenarios where individuals may not have a fixed daily commute destination, by separating travelers into 2 pools--those traveling in cohorts and those traveling individually. Each of these pools travels in separate coaches, and in practice could travel in different trains at different times to avoid infection spread at stations. We implement one-off travel by earmarking a fraction of the office going individuals as one-off travelers. The one-off travelers can be effectively viewed cohorts of size one, while the remaining group is used to form cohorts of required size. We assign separate coaches for one-off travelers and normal cohorts, thereby restricting all interactions between normal cohorts and one-off travelers. Even though we don't model the transition of an individual from one-off traveler to cohort traveler, we expect this can be achieved via testing or quarantining before joining a cohort.

\section{Modeling Details}
Once the cohort assignments are done, the simulator proceeds in time steps of 6hrs. The simulator is seeded with 100 infected individuals at the start of the simulation. The start of simulation, with reference to actual timelines, is obtained as part of the simulator calibration step; we discuss in more detail in Section \ref{sec:calibration}. At time step $t$, a susceptible individual $n$ is exposed to a daily disease transmission rate $\lambda_{n}(t)$. The computation of $\lambda_{n}(t)$ takes into account the individual's interactions in all the interaction spaces they are associated with. The strength of interactions between an individual and the various interaction spaces are determined by the intervention policy active at the given time step. The intervention policies are chosen based on the announced policies by the government. Some parameters (e.g., compliance parameter, contact tracing parameters) are tuned so that the simulator output is in reasonable agreement with the actual data from the city. 

We model the contribution of an interaction space to an individual's infection rate via a mean-field approximation of the infectivity prevalent in the interaction space. Such an approximation helps us to, $(i)$ avoid simulating pair wise interactions between individuals, and $(ii)$ one common computation of the infectivity in the interaction space suffices for all individuals in the interaction space. Individual variations are modeled separately, but is easily implemented via the use of individual specific scaling factors. We follow the same methodology for cohorts too. We describe the mean-field approximations used in modeling cohorts in (\ref{eqn:lambda_intra_cohort}), (\ref{eqn:lambda_inter_cohort}) and (\ref{eqn:transmission rate trains}).

Given the disease transmission rate seen by an individual, a susceptible individual will be infected at the current timestep $t$ with probability $(1-exp(-\lambda_{n}(t) \times \Delta_{t}))$, where $\Delta_{t}$ is the duration of the simulations timestep in days (1/4 in our simulations). Thus, computation of $\lambda_{n}(t)$ for every individual at every timestep is a key component of the simulator. 

In this study, we introduce cohorts and model the contribution from trains to an individual's transmission rate. Contribution from trains is modeled in two parts: 1) contribution from the same cohort, which we call the intra-cohort interaction, and from other cohorts that share a train coach with the cohort under consideration, which we call the inter-cohort interactions. For cohort $i$, the intra-cohort transmission rate is modeled as

\begin{align}
\label{eqn:lambda_intra_cohort}
  \lambda_{intra\_cohort}(i,t) &= \beta_{coach}\tau(i)
    \sum_{n':Co(n') = i}  I_{n'}(t) \rho_{n'}(t) \kappa_{c}(n',t) ,
\end{align}
where $\tau(i)$ denotes the travel time for cohort $i$ across all journey legs, the summation is across all individuals $n'$ who belong to cohort $i$. $Co(n')$ denotes the individual to cohort mapping. $I_{n'}(t)$ denotes whether individual $n'$ is infective at time $t$, $\rho_{n'}(t)$ is the infectiousness factor for individual $n'$, and $\kappa_{c}(n',t)$ is a modulation factor for cohort related interactions for individual $n'$, and is used to model the effect of various intervention strategies. For example, when an individual $n'$'s cohort is quarantined, we set $\kappa_{c}(n',t)$ to zero, to model the individual's lack of interaction via cohorts. 

$\beta_{coach}$ denotes the transmission rate parameter, that can be used to calibrate the simulator behavior with actual observations.

The inter-cohort interaction seen by cohort $i$ is modeled as 
\begin{align}
    \label{eqn:lambda_inter_cohort}
    \lambda_{inter\_cohort}(i,t) = \sum_{j : j \ne i}  \lambda_{intra\_cohort}(j,t) \frac{OT(i,j,t)}{\tau(j)},
\end{align}
where $j$ denotes all other cohorts and $OT(i,j,t)$ denotes the overlap time in the journeys of cohorts $i$ and $j$, where two cohorts are considered to overlap only if they share a coach.

Once the intra-cohort and inter-cohort transmission rates are computed, the contribution of trains to a susceptible individual $n$'s transmission rate is modeled as
\begin{align}
   \lambda_{n}^{trains}(t) &= (\lambda_{intra\_cohort}(i,t) + \lambda_{inter\_cohort}(i,t)) \kappa_{c}(n,t),
   \label{eqn:transmission rate trains}
\end{align}
where $i$ denotes individual $n$'s cohort, and $\kappa_{c}(n,t)$ is the workplace modulation factor for individual $n$.

Unlike other interaction spaces where the interaction space is considered active for all simulation timeslots, for trains we consider them active twice daily in two 6hr simulation timeslots. The first timeslot corresponds to morning travel and the second timeslot corresponds to evening travel.

In (\ref{eqn:transmission rate trains}) we have assumed that, inside a coach, the interactions are homogeneous among all individuals in the coach. This assumption can be translated to uniform mixing of individuals in a coach. Even though we can expect cohort members to stay together and hence restrict interactions with other cohorts, the above assumption models the worst case scenario. The assumption is further reasonable as it will be difficult to implement physical distancing between cohorts and also to model the level of interactions between cohorts.

For cohorts, unlike the modeling of infection spread in other interaction spaces where we assume that the number of interactions of an individual remains constant for a given time interval, we assume that the number of interactions increases in proportion to the number of individuals traveling in a coach. We assume a linear increase in interactions as the number of individuals increase, though one could consider other alternatives, such as a monotonically increasing concave function.

\subsection{Calibration of parameters}
\label{sec:calibration}
 Like $\beta_{coach}$ there are other tunable parameters in the simulator and it is important to tune the parameters to appropriate values to obtain reasonable outputs and insights from the simulator. We perform calibration in two steps. We seed a fixed number of individuals (100 in our case) in exposed state at the start of the simulation, and simulate a \textit{no-intervention} (no mitigation strategies are enabled) scenario. We then tune the transmission parameters for home, workplace and community ($\beta_{H}, \beta_{W}, \beta_{C}$) to minimise the difference in slopes of the log of cumulative fatalities seen in the simulator to that of the actual cumulative fatalities observed in India between 26 March  2020 and 10 April 2020 (from 10 fatalities to 199 fatalities). We consider the \textit{no-intervention} policy for calibration as the fatalities before April 10 2020 can be assumed to have got infected prior to imposition of any restrictions in India. A linear fit for the log fatalities curve is based on the assumption that cumulative fatalities grow exponentially with time, which was observed to be consistent with actual data. We observe a good match in the slopes of the fatalities curves for India and Mumbai, suggesting similar disease transmission rates in Mumbai and in the whole of India in the initial days of the pandemic.
 
 Transmission parameters in interaction spaces are tied to one of the above three parameters. For example, we assume that the transmission parameter for a project team is nine times that of the transmission parameter for the larger workplace. This is based on the assumption that an individual spends 90\% of the time in office with their team members and only the remaining 10\% with the larger workplace group. We further try to equalise the contribution of household, workplace and community towards disease spread. We use {\em stochastic approximation} methods to arrive at the appropriate parameters values. Once we match the growth rate of fatalities, we calibrate the start of simulation date with the actual fatalities timeline, so that the time series of fatalities of the simulator matches in expectation with the actual data.
  \subsubsection{Calibration of $\beta_{coach}$}
 Due to sparsity of data on the impact of trains on disease transmission we have not been able to calibrate $\beta_{coach}$ independently. We use the following heuristic argument to compute a nominal $\beta_{coach}$ from $\beta_{H}$, where $\beta_{H}$ is the transmission rate parameter associated with households.
 
 The infection transmission rate seen by an individual $n$ from their household at time $t$ is modeled as 
 \begin{align}
     \lambda_{n}^H(t) &= \beta_{H}\frac{1}{n_{H}^{\alpha}} \sum_{n'=1}^{n_H} I_{n'}(t) \rho_{n'}(t) \kappa_{H}(n',t),
 \end{align}
 where the summation is across all individuals in the household, $(1-\alpha)$ denotes the crowding factor for households, $I_{n'}(t)$ denotes whether individual $n'$ is infective at time $t$, $\rho_{n'}(t)$ denotes individual $n'$'s infectiousness factor and $\kappa_{H}(n',t)$ denotes individual $n'$'s household based modulation factor.
 
 Thus, $\beta_{H}$ can be interpreted as the household transmission rate per day for an individual. Let $r_{H}$ denote the number of typical contacts for an individual at home for a day. Then, the probability of transmission from a contact can modeled as $p_{c} = \frac{\beta_{H}}{r_{H}}$.
 
   \subsubsection{Calibration of $\beta_{coach}$}
 Due to sparsity of data on the impact of trains on disease transmission we have not been able to calibrate $\beta_{coach}$ independently. We use the following heuristic argument to compute a nominal $\beta_{coach}$ from $\beta_{H}$, where $\beta_{H}$ is the transmission rate parameter associated with households.
 
 The infection transmission rate seen by an individual $n$ from their household at time $t$ is modeled as 
 \begin{align}
     \lambda_{n}^H(t) &= \beta_{H}\frac{1}{n_{H}^{\alpha}} \sum_{n'=1}^{n_H} I_{n'}(t) \rho_{n'}(t) \kappa_{H}(n',t),
     \label{eqn:lambda_home}
 \end{align}
 where the summation is across all individuals in the household, $(1-\alpha)$ denotes the crowding factor for households, $I_{n'}(t)$ denotes whether individual $n'$ is infective at time $t$, $\rho_{n'}(t)$ denotes individual $n'$'s infectiousness factor and $\kappa_{H}(n',t)$ denotes individual $n'$'s household based modulation factor.
 
 Thus, $\beta_{H}$ can be interpreted as the household transmission rate per day for an individual. Let $r_{H}$ denote the number of typical contacts for an individual at home for a day. Then, the probability of transmission from a contact can modeled as $p_{c} = \frac{\beta_{H}}{r_{H}}$.
 
 Let the number of typical contacts per minute per individual in a train coach be $r_T$. Then, the effective transmission rate per minute per individual in coach, $\hat{\beta}_{coach}$, can be expressed in terms of $r_T$ as 

\begin{align}
     \hat{\beta}_{coach} =  p_c r_{T} = \frac{\beta_{H} r_{T}}{r_{H}}.
\end{align}
 
If we assume that the number of close contacts per day in a household $r_{H} = 50$ contacts per day, and close contacts per minute per individual in a coach $r_{T} = 1 / 100$, $\hat{\beta}_{coach}$ can be computed in terms of $\beta_{H}$ as:

 \begin{align}
     \hat{\beta}_{coach} = \beta_{H} \frac{r_{T}}{r_{H}}
    =\beta_{H} \frac{1} {100} \frac{1}{50}
    =\beta_{H} 0.0002.
 \end{align}

In the simulator, for every simulation timestep $\Delta_{t}$, the transmission rate is further multiplied by $\Delta_{t}$ to obtain the mean disease transmitting contacts per simulation timestep. For trains, since we already account for commute time, and since we assume that one journey is restricted to one simulation time step, we need to discount the further multiplication by $\Delta_{t}$. Thus, a nominal value for the $\beta_{coach}$ parameter we use in simulator, in terms of $\beta_{H}$ can be obtained as

 \begin{align}
    \beta_{coach}= \frac{\hat{\beta}_{coach}}{\Delta_{t}}
    = \hat{\beta}_{coach} 4
    = \beta_{H} 0.0008,
 \end{align}
 where we have used $\Delta_{t} = 1/4$ days, to account for a simulation timestep duration of 6hrs.

\subsection{Intervention Modeling}
A key feature of the simulator in \cite{ref:J-IISc} is its ability to simulate various time varying  intervention strategies. Interventions are modeled by modulating an individual's edge weights with various interaction space. For example, when an individual is self-isolated at home, contact rates with their household is reduced by 25\%, contact rates with their workplace is reduced to zero and contact rates with the community is reduced to 10\%. The simulator also supports testing and contact tracing protocols. Contact tracing in the close network of an individual can be initiated for each of the following events: ($i$) an individual is hospitalised, ($ii$) an individual tests positive, ($iii$) an individual reports symptoms. The fraction of such events that trigger contact tracing and the fraction of individuals who would be contact traced are all configurable.

\subsection{Intervention Modeling - Cohorts}

In the specific case of cohorts, we study the impact of isolating an entire cohort when an individual is hospitalised or tested positive or is sufficiently symptomatic. When an individual is hospitalised or tested positive, then all their cohort members are placed under self isolation. A symptomatic individual may self-declare or be detected at a station and that can also trigger isolation of the other cohort members.

\begin{table}[t]
\footnotesize	
\caption{Model parameters}
\label{tab:params}
\centering
\begin{tabular}{lccc}
 \toprule
 \textit{Parameter} & \textit{Symbol} & \textit{Mumbai} \\
 \midrule
 Transmission coefficient at home   & $\beta_h$ &  0.7928 (calibrated)\\
 Transmission coefficient at school & $\beta_s$ & 0.2834 (calibrated) \\
 Transmission coefficient at workplace & $\beta_w$ & 0.1417 (calibrated)\\
 Transmission coefficient at community & $\beta_c$ &  0.0149 (calibrated)\\
 Subnetwork upscale factor & $\tilde{\beta}$ & 9 \\
 Transmission coefficient at transport space & $\beta_{coach}$ & 0.0005\\
 Household crowding & $1 - \alpha$ & 0.2\\
 Community crowding & $r_c$ & 2\\
 Distance kernel $f(d) = 1/(1+(d/a)^b)$ & $(a,b)$ &  $(2.709,1.279)$\\
 Infectiousness shape (Gamma distributed) & (shape,scale) &  $(0.25, 4)$ \\
 Severity probability & $\Pr\{C_n = 1\}$ &  0.5\\
 Age stratification & $M_{n,n'}$ & Not used\\
 Project subnetwork size range & $n_{\mathscr{W}(n)}$ &  $3-10$ \\
 Family friends' subnetwork range & no symbol & 2-5 families \\
 \bottomrule
\end{tabular}

\end{table}
\section{Results and key findings \label{sec:results}}
\begin{figure*}
 \centering
   \begin{subfigure}[b]{0.3\textwidth}
    \includegraphics[width=\textwidth]{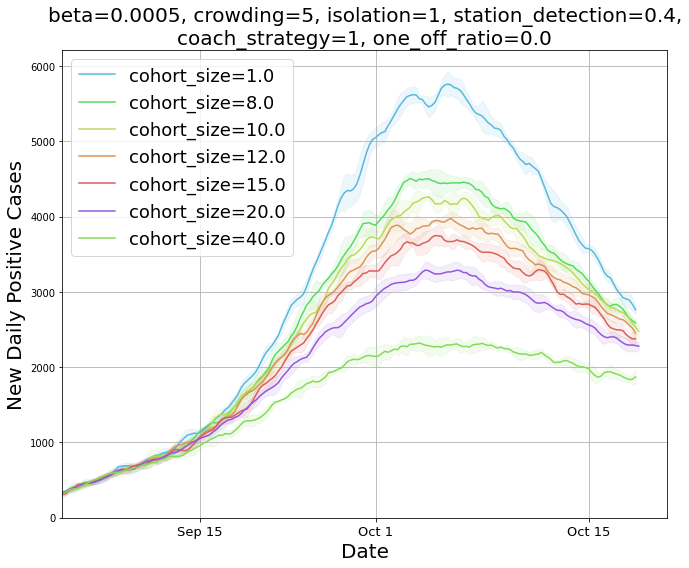}
    \caption{Cohort size}
    \label{ref:figure_cohort_size}
  \end{subfigure}
  \hfill
  \begin{subfigure}[b]{0.3\textwidth}
    \includegraphics[width=\textwidth]{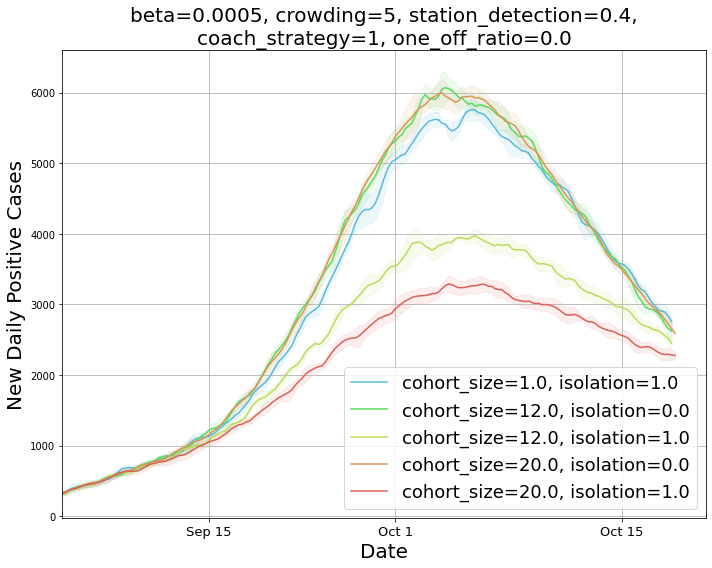}
    \caption{Cohort intervention policy }
    \label{ref:figure_isolation}
  \end{subfigure}
  \hfill
  \begin{subfigure}[b]{0.3\textwidth}
    \includegraphics[width=\textwidth]{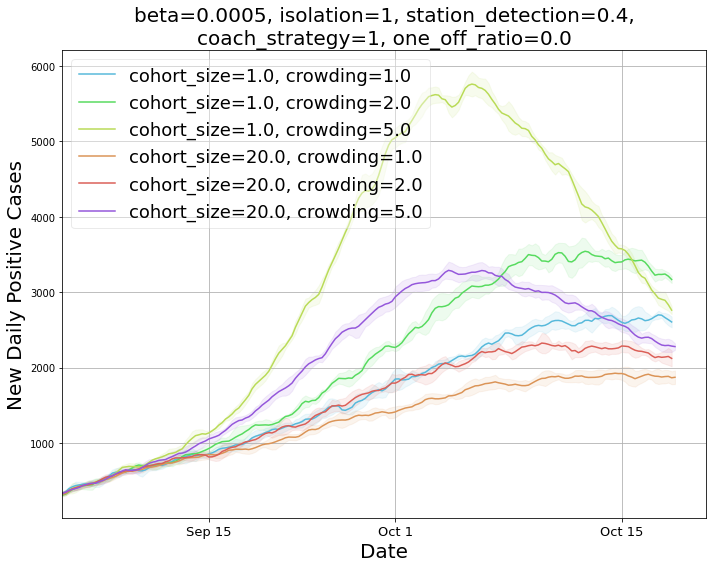}
    \caption{Crowding}
    \label{ref:figure_crowding}
  \end{subfigure}
  \hfill
  \begin{subfigure}[b]{0.3\textwidth}
    \includegraphics[width=\textwidth]{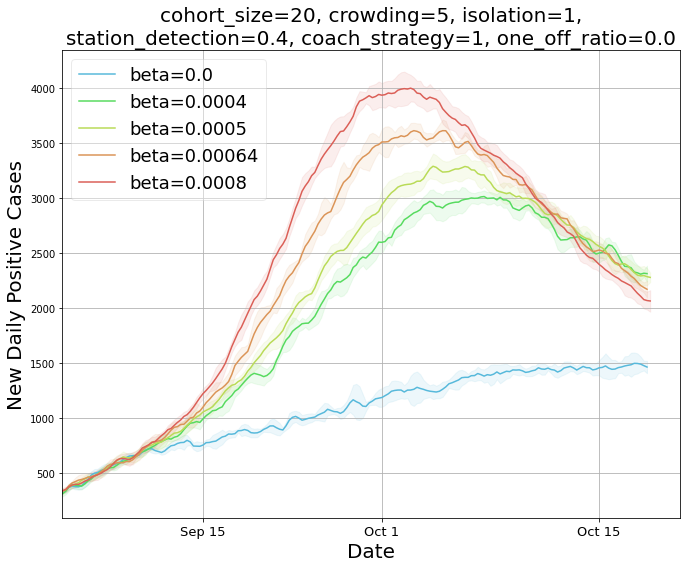}
    \caption{$\beta_{coach}$}
    \label{ref:figure_beta}
  \end{subfigure}
\hfill
\begin{subfigure}[b]{0.3\textwidth}
    \includegraphics[width=\textwidth]{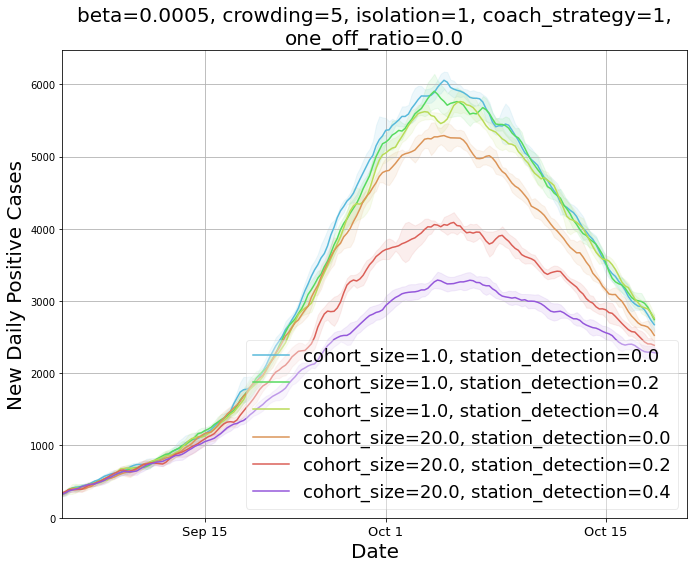}
    \caption{Symptomatic detection at station}
    \label{ref:figure_symptomatic_threshold}
  \end{subfigure}
\hfill
  \begin{subfigure}[b]{0.3\textwidth}
    \includegraphics[width=\textwidth]{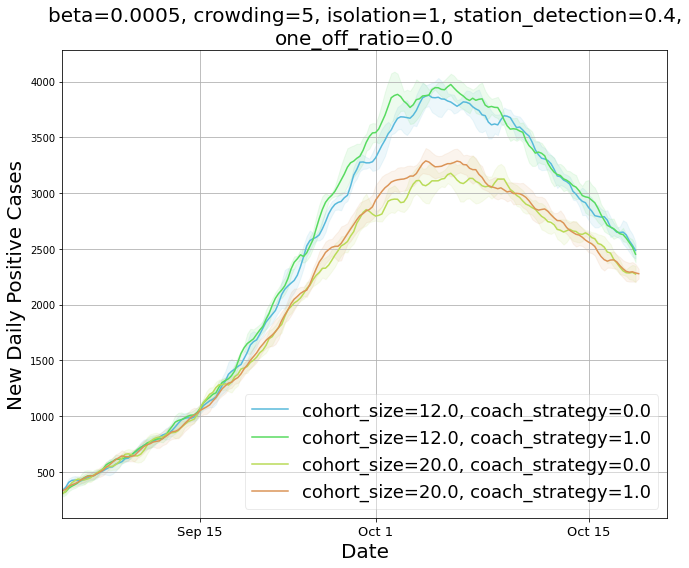}
    \caption{Cohort-to-coach mapping strategy}
    \label{ref:figure_cohort_strategy}
  \end{subfigure}
\hfill
\caption{
Y-axis represents daily new detected positive cases. Parameters which vary are labeled in the legend, where as parameters which don't are noted above the plot. The shaded region represents 1 standard deviation away from the mean, estimated over 5 simulation runs with the selected configuration.
Beta represents $\beta_{coach}$, the transmission rate parameter of the interaction in train coaches. 
Crowding represents the crowding factor of train coaches, which impacts occupancy limit of a train coach.
Isolation value of 0 or 1 represents lack  or presence of cohort isolation policy respectively. 
Coach\_strategy of 0 or 1 represents static or dynamic coach assignment respectively. 
One\_off\_ratio represents proportion of travelers that travel one off, and the remainder $(1.0 - one\_off\_ratio)$ travel in cohorts of the selected cohort\_size.
Station detection represents the proportion of symptomatic infected commuters detected at the station, via thermal screening or other testing mechanisms.
}
\label{ref:daily_plots_group}
\end{figure*}

Based on the methodology described in Section~\ref{ref:Cohorts}, 3.74M individuals take trains for daily commute in our simulations. In terms of interventions, we consider a pre-lockdown period starting from March 16 and lockdown extending till May 18 in Mumbai. From May 18, the following interventions are active: $(i)$ 14 day home quarantining of symptomatic individuals and their household members, for compliant households, $(ii)$ social distancing by compliant elderly (age > 65), $(iii)$ schools are closed. Further, post May 18, we assume a phased re-opening of offices, with 5\% offices operating till May 31, followed by 15\% attendance till June 30, followed by 25\% attendance till July 31, followed by 33\% attendance till Aug 31 and full attendance thereafter. Further, we assume masks were enforced from April 09. Compliance probabilities to the regulations (including mask wearing) are 60\% in high density areas and 40\% in other areas. Testing and contact tracing is active from March 16. On hospitalisation or identification of a positive case, contact tracing, isolation and testing is triggered, based on a set of specified probabilities. This models the {\em test-isolate} intervention. Trains are assumed to start from September 07. Progressive containment of wards based on number of hospitalised cases is active till the end of August.

We run the simulation from mid February, but exclude presenting results for the duration prior to the anticipated restart date for the \textit{locals}. As an implementation detail, we built in our model the ability to store and load state at any timestep, and were able to store state for state prior to the anticipated \textit{locals} restart date. We only stored the infection state of individuals. Randomization seed, and other parameters related to cohorting were not stored. This ability to store once and load multiple times helped reduce computation time significantly. In a few scenarios for long term projections and estimating network saturation, we run the simulator for significantly longer duration. 

\begin{figure}
  \begin{subfigure}[t]{0.23\textwidth}
    \includegraphics[width=\textwidth]{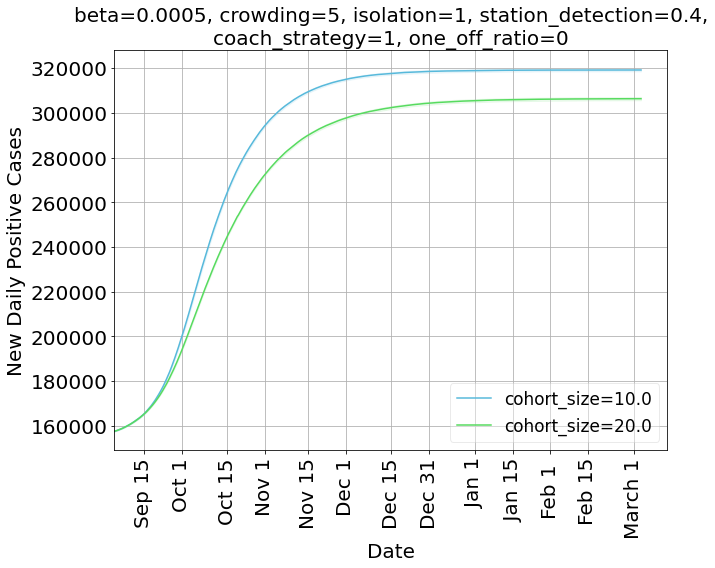}
    \caption{Cumulative positive cases}
    \label{ref:fig_long_cum}
  \end{subfigure}
\hfill
  \begin{subfigure}[t]{0.23\textwidth}
    \includegraphics[width=\textwidth]{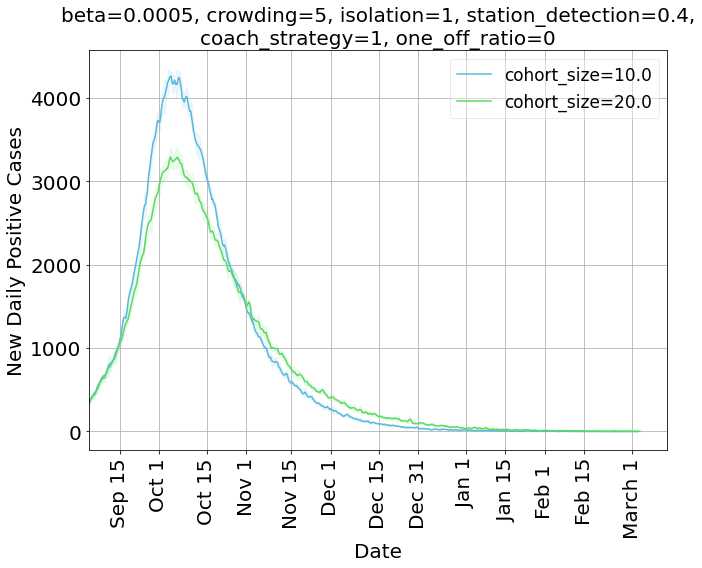}
    \caption{Daily positive cases}
    \label{ref:fig_long_daily}
  \end{subfigure}
\hfill
\caption{
Effect of cohort size on peak daily case load and total case load at saturation.}
\label{ref:fig_cum_daily}.
\end{figure}

In Figure \ref{ref:daily_plots_group}, we plot new daily positive cases, which represent the new cases detected on the specified date. This value is much lower than the actual underlying new infections on that day. All plots in the group peak and then trend downwards. To explain this, we plot Figure \ref{ref:fig_cum_daily} which shows that the cumulative detected cases to that date side by side with new daily positive cases, and we observe that the network saturates by Dec, with inflection point of new daily positive cases in mid Oct, which matches with the peak of new daily cases. Also interestingly we see that network saturation level for larger cohort sizes is much lower, which shows that cohorting helps reduce total cases significantly.

It is important to note that the daily positive cases take into account interaction in other interaction spaces (household, office, neighbourhood, community) in addition to the transport interaction space. The plots thus show the impact of cohorting strategies on the progression of the \emph{overall disease burden in the city}. Another feature of the simulator is that it models spatial variance in population density and the associated increased contacts in slum areas, and therefore increased spread in these areas. A third important feature worth highlighting is that we model contact tracing, which help contain the spread, this helps to not only match actual observed cases, but also model the benefit of contact tracing enabled case isolation in a realistic way.

\subsection{Details}

\medskip
\noindent
\textbf{1. Impact of cohort size.}\ 
Figures~\ref{ref:figure_cohort_size} and  \ref{ref:figure_isolation} show the change in infection spread dynamics for various cohort sizes. We observe:
\begin{itemize}
    \item Without isolation, all cohorts have similar dynamics.
    \item With isolation, we observe reduction in disease spread, as potentially asymptomatic cases are removed from the mix. 
    \item Significant reduction in disease spread is observed as cohort sizes increases.
\end{itemize}


\medskip
\noindent
\textbf{2. Impact of crowding factor.}\  
Figure \ref{ref:figure_crowding} plots daily positive cases for different crowding factors. 
\begin{itemize}
    \item Disease spread is extremely sensitive to crowding. 
    \item Crowding factor of $1$ without isolation has lower daily cases than crowding factor $2$ with isolation, suggesting that reducing the crowding in trains by half has more impact than enforcing isolation of cohorts. 
\end{itemize}

\medskip
\noindent
\textbf{3. Impact of contact rate.}\  
In Figure \ref{ref:figure_beta}, we plot daily positive cases for different $\beta_{coach}$, the contact rate parameter for cohorts. Since we do not have a calibrated $\beta$ value, it is important to study the robustness of the results to various $\beta$ values. We observe:
\begin{itemize}
    \item Higher $\beta_{coach}$ values cause higher disease transmission.
    \item Higher $\beta_{coach}$ values reach the peak earlier.
    \item We observe similar trend in the disease progression curves for different $\beta$ values, suggesting that the qualitative results are robust to variations in $\beta$ values.
\end{itemize}

\medskip
\noindent
\textbf{4. Impact of detection probability.}\ 
In Figure~\ref{ref:figure_symptomatic_threshold}, we study the impact of varying the detection probability of symptomatic individuals at stations. A detection triggers the isolation of the entire cohort. Detection of positive symptomatic individuals is critical for cohorting to succeed and could potentially be achieved using thermal scanners at stations, and/or by employing random testing of individuals at stations. As detection increases, spread of diseases falls, and without detection at cohorting has only marginal improvement compared to no-cohorting scenarios.

 \medskip
\noindent
\textbf{5. Impact of inter-day cohort-to-coach assignment strategies.}\ 
 In Figure \ref{ref:figure_cohort_strategy}, we study the impact of static coach assignment, where restricting cohorts to travel with the same set of cohorts on a daily basis against a much relaxed policy of allowing cohorts choose their train of choice and time of travel. The only restriction is that the cohorts should travel together. We observe no significant benefits from static coach assignment, suggesting strict coach assignment may not be warranted. 
 
\begin{figure}
  \begin{subfigure}[t]{0.23\textwidth}
        \includegraphics[width=\textwidth]{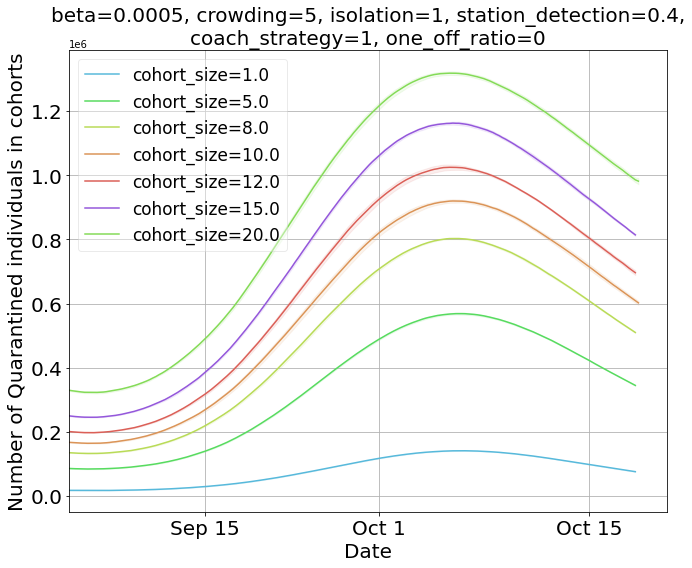}
        \caption{\# of individuals quarantined due to cohorting in locals}
        \label{ref:figure_daily_cohort_quarantined}
  \end{subfigure}
\hfill
  \begin{subfigure}[t]{0.23\textwidth}
        \includegraphics[width=\textwidth]{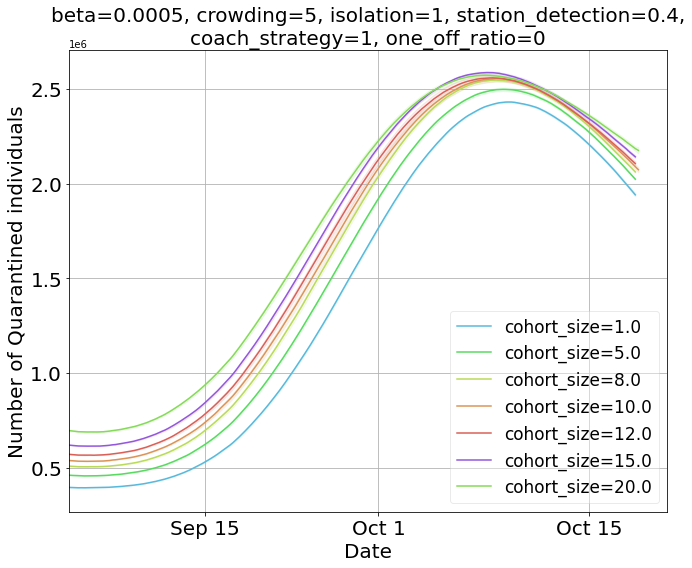}
        \caption{\# of total individuals quarantined in the city}
        \label{ref:figure_daily_quarantined}
  \end{subfigure}
\hfill
\caption{
As cohort size increases, there is a greater contribution of quarantined cases due to cohorting, but the increase in total quarantined cases in the city is marginal.}
\label{quarantine_plots_group}
\end{figure}

 \medskip
\noindent
\textbf{6. Impact on count of quarantined individuals}\ 
 \ref{ref:figure_daily_cohort_quarantined} shows that the number of people quarantined due to cohorting increases with increased cohort sizes.
 Interestingly, the increase in total number of quarantined individuals in city due to cohorting is small (Figure~\ref{ref:figure_daily_quarantined}). 
 This can be attributed to the observation that cohorting with reasonable cohort size can reduce the daily positive cases substantially, and stem disease progression (Figure \ref{ref:figure_cohort_size}).
 Cohorting is able to quarantine individuals even before testing, and contact tracing mechanisms kick in, and has an ability to smartly identify \& isolate.
 
\begin{figure}
  \centering
    \begin{subfigure}[t]{0.23\textwidth}
    \includegraphics[width=\textwidth]{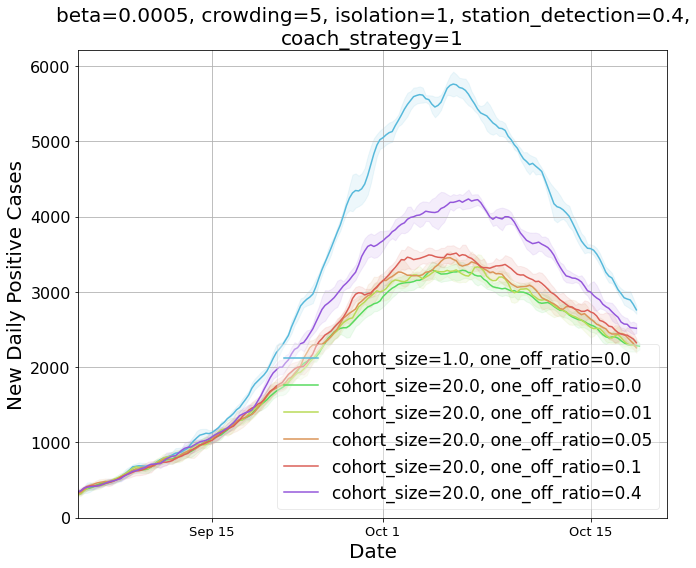}
  \end{subfigure}
\hfill
  \begin{subfigure}[t]{0.23\textwidth}
    \includegraphics[width=\textwidth]{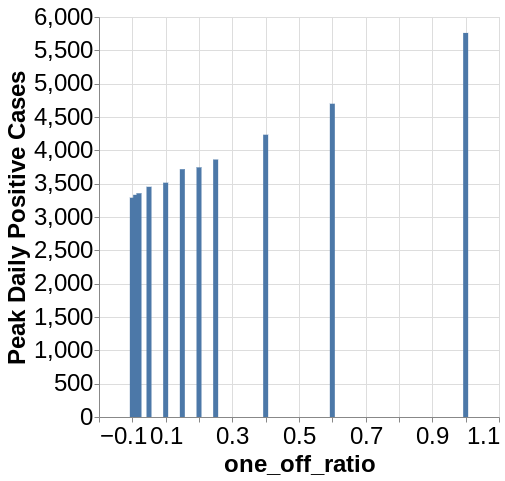}
  \end{subfigure}
 \caption{Effect of one-off travelers: As fraction of one-off travelers increase, the daily positive cases increase but partial cohorting does far better than business as usual (cohort size of 1).}
 \label{ref:one_off_plot_group}
\end{figure}

 \medskip
\noindent
\textbf{7. Impact of one-off travel.}\ 
Figure \ref{ref:one_off_plot_group} shows the impact of allowing one-off travel along side cohorts. We assume one-off travelers use separate coaches and thus avoid interacting with those in cohorts. 
Figure \ref{ref:one_off_plot_group} (right) shows new daily positive cases detected with increasing cohort sizes. 
We observe that even with significant proportion of one-off travelers (40\%), cohorting strategy reduces disease transmission significantly.

\subsection{Ideal Cohort Size}
Even though our study finds larger cohort sizes attenuate disease transmission more severely, there are practical considerations like coach capacity utilization, and enforceability of isolation policies. In practice a cohort size of 12 to 20 might be more appealing.

\section{Key findings and policy implications}
\begin{itemize}
    \item Cohorting can significantly reduce disease transmission. Larger cohort sizes are more effective at reducing disease transmission.
    \item Effectiveness of cohorting depends on effectiveness of timely case detection (via thermal screening for symptomatic cases at stations) and isolation (enforcing quarantine for all cohort members when a single member is detected to be positive). 
    \item Static coach assignment does not have significant impact on reducing disease transmission as compared to dynamic coach assignment.
    \item Disease transmission is most sensitive to crowding in trains, and crowding needs to be effectively managed. While reducing crowding on train coaches, care must be taken to not crowd the stations. This aspect needs to be studied further.
    \item One-off travel up to 10\% will only show marginal increase in disease transmission, and even at 40\% one-off travel, disease transmission can be significantly reduced by cohorting. There is benefit in incremental implementation of cohorting. 
    \item While the model has focused on Mumbai metro rail system, we believe these findings are generally applicable to other public transit systems (like metros, buses) in India as well other large metropolitan areas.
\end{itemize}

\begin{acks}
This work was supported by a Google Gift Grant. The SahasraT supercomputing cluster at the Supercomputer Education and Research Centre (SERC), Indian Institute of Science, was used for most of the simulations. NKV and SS's work was done at the IISc-Cisco Centre for Networked Intelligence; the centre's support is gratefully acknowledged. RS's work was done while on sabbatical leave at Strand Life Sciences; their support is gratefully acknowledged. NKV, SS and RS are grateful for the discussions, comments, suggestions from Sandeep Juneja, Ramprasad Saptharishi, Piyush Srivastava and Prahladh Harsha from the School of Technology and Computer Science, Tata Institute of Fundamental Research, Mumbai. 

We (JC, SV, LW, AA, MV) thank members of the Biocomplexity Institute and Initiative, University of Virginia for useful discussion and suggestions.  This work was partially supported by National Institutes of Health (NIH) Grant R01GM109718, NSF BIG DATA Grant IIS-1633028, NSF DIBBS Grant OAC-1443054,
NSF Grant No.: OAC-1916805, NSF Expeditions in Computing Grant CCF-1918656, CCF-1917819 \& NSF RAPID CNS-2028004, NSF RAPID OAC-2027541.  Any opinions, findings, and conclusions or recommendations expressed in this material are those of the author(s) and do not necessarily reflect the views of the funding agencies.
\end{acks}

\balance
\bibliographystyle{ACM-Reference-Format} 
\bibliography{references}


\begin{thebibliography}{35}


\ifx \showCODEN    \undefined \def \showCODEN     #1{\unskip}     \fi
\ifx \showDOI      \undefined \def \showDOI       #1{#1}\fi
\ifx \showISBNx    \undefined \def \showISBNx     #1{\unskip}     \fi
\ifx \showISBNxiii \undefined \def \showISBNxiii  #1{\unskip}     \fi
\ifx \showISSN     \undefined \def \showISSN      #1{\unskip}     \fi
\ifx \showLCCN     \undefined \def \showLCCN      #1{\unskip}     \fi
\ifx \shownote     \undefined \def \shownote      #1{#1}          \fi
\ifx \showarticletitle \undefined \def \showarticletitle #1{#1}   \fi
\ifx \showURL      \undefined \def \showURL       {\relax}        \fi
\providecommand\bibfield[2]{#2}
\providecommand\bibinfo[2]{#2}
\providecommand\natexlab[1]{#1}
\providecommand\showeprint[2][]{arXiv:#2}

\bibitem[\protect\citeauthoryear{Adiga, Chu, Eubank, Kuhlman, Lewis, Marathe,
  Marathe, Nordberg, Swarup, Vullikanti, et~al\mbox{.}}{Adiga
  et~al\mbox{.}}{2018}]%
        {adiga2018disparities}
\bibfield{author}{\bibinfo{person}{Abhijin Adiga}, \bibinfo{person}{Shuyu Chu},
  \bibinfo{person}{Stephen Eubank}, \bibinfo{person}{Christopher~J Kuhlman},
  \bibinfo{person}{Bryan Lewis}, \bibinfo{person}{Achla Marathe},
  \bibinfo{person}{Madhav Marathe}, \bibinfo{person}{Eric~K Nordberg},
  \bibinfo{person}{Samarth Swarup}, \bibinfo{person}{Anil Vullikanti},
  {et~al\mbox{.}}} \bibinfo{year}{2018}\natexlab{}.
\newblock \showarticletitle{Disparities in spread and control of influenza in
  slums of Delhi: findings from an agent-based modelling study}.
\newblock \bibinfo{journal}{\emph{BMJ open}} \bibinfo{volume}{8},
  \bibinfo{number}{1} (\bibinfo{year}{2018}).
\newblock


\bibitem[\protect\citeauthoryear{Adiga, Wang, Sadilek, Tendulkar,
  Venkatramanan, Vullikanti, Aggarwal, Talekar, Ben, Chen, et~al\mbox{.}}{Adiga
  et~al\mbox{.}}{2020}]%
        {adiga2020interplay}
\bibfield{author}{\bibinfo{person}{Aniruddha Adiga}, \bibinfo{person}{Lijing
  Wang}, \bibinfo{person}{Adam Sadilek}, \bibinfo{person}{Ashish Tendulkar},
  \bibinfo{person}{Srinivasan Venkatramanan}, \bibinfo{person}{Anil
  Vullikanti}, \bibinfo{person}{Gaurav Aggarwal}, \bibinfo{person}{Alok
  Talekar}, \bibinfo{person}{Xue Ben}, \bibinfo{person}{Jiangzhuo Chen},
  {et~al\mbox{.}}} \bibinfo{year}{2020}\natexlab{}.
\newblock \showarticletitle{Interplay of global multi-scale human mobility,
  social distancing, government interventions, and COVID-19 dynamics}.
\newblock \bibinfo{journal}{\emph{medRxiv}} (\bibinfo{year}{2020}).
\newblock


\bibitem[\protect\citeauthoryear{Antelmi, Cordasco, Spagnuolo, and
  Scarano}{Antelmi et~al\mbox{.}}{2020}]%
        {antelmi2020design}
\bibfield{author}{\bibinfo{person}{Alessia Antelmi}, \bibinfo{person}{Gennaro
  Cordasco}, \bibinfo{person}{Carmine Spagnuolo}, {and}
  \bibinfo{person}{Vittorio Scarano}.} \bibinfo{year}{2020}\natexlab{}.
\newblock \showarticletitle{A Design-Methodology for Epidemic Dynamics via
  Time-Varying Hypergraphs}. In \bibinfo{booktitle}{\emph{Proceedings of the
  19th International Conference on Autonomous Agents and MultiAgent Systems
  (AAMAS' 20)}}. \bibinfo{pages}{61--69}.
\newblock


\bibitem[\protect\citeauthoryear{Baker, Basu, Cropper, Lall, and
  Takeuchi}{Baker et~al\mbox{.}}{2005}]%
        {ref:mumbai_commute_world_bank_report}
\bibfield{author}{\bibinfo{person}{Judy Baker}, \bibinfo{person}{Rakhi Basu},
  \bibinfo{person}{Maureen Cropper}, \bibinfo{person}{Somik Lall}, {and}
  \bibinfo{person}{Akie Takeuchi}.} \bibinfo{year}{2005}\natexlab{}.
\newblock \bibinfo{booktitle}{\emph{Urban poverty and transport: the case of
  Mumbai}}.
\newblock \bibinfo{publisher}{The World Bank}.
\newblock


\bibitem[\protect\citeauthoryear{Balmer, Meister, Rieser, Nagel, and
  Axhausen}{Balmer et~al\mbox{.}}{2008}]%
        {balmer2008agent}
\bibfield{author}{\bibinfo{person}{Michael Balmer}, \bibinfo{person}{Konrad
  Meister}, \bibinfo{person}{Marcel Rieser}, \bibinfo{person}{Kai Nagel}, {and}
  \bibinfo{person}{Kay~W Axhausen}.} \bibinfo{year}{2008}\natexlab{}.
\newblock \showarticletitle{Agent-based simulation of travel demand: Structure
  and computational performance of MATSim-T}.
\newblock \bibinfo{journal}{\emph{Arbeitsberichte Verkehrs-und Raumplanung}}
  \bibinfo{volume}{504} (\bibinfo{year}{2008}).
\newblock


\bibitem[\protect\citeauthoryear{Barrett, Bisset, Eubank, Feng, and
  Marathe}{Barrett et~al\mbox{.}}{2008}]%
        {barrett2008episimdemics}
\bibfield{author}{\bibinfo{person}{Christopher~L Barrett},
  \bibinfo{person}{Keith~R Bisset}, \bibinfo{person}{Stephen~G Eubank},
  \bibinfo{person}{Xizhou Feng}, {and} \bibinfo{person}{Madhav~V Marathe}.}
  \bibinfo{year}{2008}\natexlab{}.
\newblock \showarticletitle{EpiSimdemics: an efficient algorithm for simulating
  the spread of infectious disease over large realistic social networks}. In
  \bibinfo{booktitle}{\emph{SC'08: Proceedings of the 2008 ACM/IEEE Conference
  on Supercomputing}}. IEEE, \bibinfo{pages}{1--12}.
\newblock


\bibitem[\protect\citeauthoryear{Block, Hoffman, Raabe, Dowd, Rahal, Kashyap,
  and Mills}{Block et~al\mbox{.}}{2020}]%
        {block2020social}
\bibfield{author}{\bibinfo{person}{Per Block}, \bibinfo{person}{Marion
  Hoffman}, \bibinfo{person}{Isabel~J Raabe}, \bibinfo{person}{Jennifer~Beam
  Dowd}, \bibinfo{person}{Charles Rahal}, \bibinfo{person}{Ridhi Kashyap},
  {and} \bibinfo{person}{Melinda~C Mills}.} \bibinfo{year}{2020}\natexlab{}.
\newblock \showarticletitle{Social network-based distancing strategies to
  flatten the COVID-19 curve in a post-lockdown world}.
\newblock \bibinfo{journal}{\emph{Nature Human Behaviour}}
  (\bibinfo{year}{2020}), \bibinfo{pages}{1--9}.
\newblock


\bibitem[\protect\citeauthoryear{Boscoe, Henry, and Zdeb}{Boscoe
  et~al\mbox{.}}{2012}]%
        {distanceaFactor}
\bibfield{author}{\bibinfo{person}{Francis~P. Boscoe},
  \bibinfo{person}{Kevin~A. Henry}, {and} \bibinfo{person}{Michael~S. Zdeb}.}
  \bibinfo{year}{2012}\natexlab{}.
\newblock \showarticletitle{A Nationwide Comparison of Driving Distance Versus
  Straight-Line Distance to Hospitals}.
\newblock \bibinfo{journal}{\emph{The Professional Geographer}}
  \bibinfo{volume}{64}, \bibinfo{number}{2} (\bibinfo{year}{2012}),
  \bibinfo{pages}{188--196}.
\newblock
\urldef\tempurl%
\url{https://doi.org/10.1080/00330124.2011.583586}
\showDOI{\tempurl}
\showeprint{https://doi.org/10.1080/00330124.2011.583586}


\bibitem[\protect\citeauthoryear{Chen, Chu, Chungbaek, Khan, Kuhlman, Marathe,
  Mortveit, Vullikanti, and Xie}{Chen et~al\mbox{.}}{2016}]%
        {chen2016effect}
\bibfield{author}{\bibinfo{person}{Jiangzhuo Chen}, \bibinfo{person}{Shuyu
  Chu}, \bibinfo{person}{Youngyun Chungbaek}, \bibinfo{person}{Maleq Khan},
  \bibinfo{person}{Christopher Kuhlman}, \bibinfo{person}{Achla Marathe},
  \bibinfo{person}{Henning Mortveit}, \bibinfo{person}{Anil Vullikanti}, {and}
  \bibinfo{person}{Dawen Xie}.} \bibinfo{year}{2016}\natexlab{}.
\newblock \showarticletitle{Effect of modelling slum populations on influenza
  spread in Delhi}.
\newblock \bibinfo{journal}{\emph{BMJ open}} \bibinfo{volume}{6},
  \bibinfo{number}{9} (\bibinfo{year}{2016}).
\newblock


\bibitem[\protect\citeauthoryear{Cooley, Brown, Cajka, Chasteen, Ganapathi,
  Grefenstette, Hollingsworth, Lee, Levine, Wheaton, et~al\mbox{.}}{Cooley
  et~al\mbox{.}}{2011}]%
        {cooley2011role}
\bibfield{author}{\bibinfo{person}{Philip Cooley}, \bibinfo{person}{Shawn
  Brown}, \bibinfo{person}{James Cajka}, \bibinfo{person}{Bernadette Chasteen},
  \bibinfo{person}{Laxminarayana Ganapathi}, \bibinfo{person}{John
  Grefenstette}, \bibinfo{person}{Craig~R Hollingsworth},
  \bibinfo{person}{Bruce~Y Lee}, \bibinfo{person}{Burton Levine},
  \bibinfo{person}{William~D Wheaton}, {et~al\mbox{.}}}
  \bibinfo{year}{2011}\natexlab{}.
\newblock \showarticletitle{The role of subway travel in an influenza epidemic:
  a New York City simulation}.
\newblock \bibinfo{journal}{\emph{Journal of urban health}}
  \bibinfo{volume}{88}, \bibinfo{number}{5} (\bibinfo{year}{2011}),
  \bibinfo{pages}{982}.
\newblock


\bibitem[\protect\citeauthoryear{Del~Valle, Stroud, Smith, Mniszewski, Riese,
  Sydoriak, and Kubicek}{Del~Valle et~al\mbox{.}}{2006}]%
        {del2006episims}
\bibfield{author}{\bibinfo{person}{Sara~Y Del~Valle},
  \bibinfo{person}{Phillip~D Stroud}, \bibinfo{person}{James~P Smith},
  \bibinfo{person}{Susan~M Mniszewski}, \bibinfo{person}{Jane~M Riese},
  \bibinfo{person}{Stephen~J Sydoriak}, {and} \bibinfo{person}{Deborah~A
  Kubicek}.} \bibinfo{year}{2006}\natexlab{}.
\newblock \showarticletitle{EpiSimS: epidemic simulation system}.
\newblock \bibinfo{journal}{\emph{Los Alamos, NM: Los Alamos National
  Laboratory}} (\bibinfo{year}{2006}).
\newblock


\bibitem[\protect\citeauthoryear{et.al.}{et.al.}{2020}]%
        {ref:J-IISc}
\bibfield{author}{\bibinfo{person}{Shubhada~Agrawal et.al.}}
  \bibinfo{year}{2020}\natexlab{}.
\newblock \showarticletitle{City-Scale Agent-Based Simulators for the Study of
  Non-Pharmaceutical Interventions in the Context of the COVID-19 Epidemic}.
\newblock \bibinfo{journal}{\emph{Journal of the Indian Institute of Science}}
  \bibinfo{volume}{100}, \bibinfo{number}{4} (\bibinfo{year}{2020}),
  \bibinfo{pages}{809--847}.
\newblock
\showISSN{0019-4964}


\bibitem[\protect\citeauthoryear{Eubank, Guclu, Kumar, Marathe, Srinivasan,
  Toroczkai, and Wang}{Eubank et~al\mbox{.}}{2004}]%
        {eubank2004modelling}
\bibfield{author}{\bibinfo{person}{Stephen Eubank}, \bibinfo{person}{Hasan
  Guclu}, \bibinfo{person}{VS~Anil Kumar}, \bibinfo{person}{Madhav~V Marathe},
  \bibinfo{person}{Aravind Srinivasan}, \bibinfo{person}{Zoltan Toroczkai},
  {and} \bibinfo{person}{Nan Wang}.} \bibinfo{year}{2004}\natexlab{}.
\newblock \showarticletitle{Modelling disease outbreaks in realistic urban
  social networks}.
\newblock \bibinfo{journal}{\emph{Nature}} \bibinfo{volume}{429},
  \bibinfo{number}{6988} (\bibinfo{year}{2004}), \bibinfo{pages}{180--184}.
\newblock


\bibitem[\protect\citeauthoryear{Glaeser, Gorback, and Redding}{Glaeser
  et~al\mbox{.}}{2020}]%
        {glaeser2020princeton}
\bibfield{author}{\bibinfo{person}{Edward~L Glaeser},
  \bibinfo{person}{Caitlin~S Gorback}, {and} \bibinfo{person}{Stephen~J
  Redding}.} \bibinfo{year}{2020}\natexlab{}.
\newblock \bibinfo{booktitle}{\emph{How much does covid-19 increase with
  mobility? evidence from new york and four other us cities}}.
\newblock \bibinfo{type}{{T}echnical {R}eport}. \bibinfo{institution}{National
  Bureau of Economic Research}.
\newblock


\bibitem[\protect\citeauthoryear{Gosc{\'e}, Barton, and Johansson}{Gosc{\'e}
  et~al\mbox{.}}{2014}]%
        {gosce2014corridor}
\bibfield{author}{\bibinfo{person}{Lara Gosc{\'e}}, \bibinfo{person}{David~AW
  Barton}, {and} \bibinfo{person}{Anders Johansson}.}
  \bibinfo{year}{2014}\natexlab{}.
\newblock \showarticletitle{Analytical modelling of the spread of disease in
  confined and crowded spaces}.
\newblock \bibinfo{journal}{\emph{Scientific reports}}  \bibinfo{volume}{4}
  (\bibinfo{year}{2014}), \bibinfo{pages}{4856}.
\newblock


\bibitem[\protect\citeauthoryear{Gosc{\'e} and Johansson}{Gosc{\'e} and
  Johansson}{2018}]%
        {gosce2018londonTube}
\bibfield{author}{\bibinfo{person}{Lara Gosc{\'e}} {and}
  \bibinfo{person}{Anders Johansson}.} \bibinfo{year}{2018}\natexlab{}.
\newblock \showarticletitle{Analysing the link between public transport use and
  airborne transmission: mobility and contagion in the London underground}.
\newblock \bibinfo{journal}{\emph{Environmental Health}} \bibinfo{volume}{17},
  \bibinfo{number}{1} (\bibinfo{year}{2018}), \bibinfo{pages}{84}.
\newblock


\bibitem[\protect\citeauthoryear{Hackl and Dubernet}{Hackl and
  Dubernet}{2019}]%
        {hackl2019epidemic}
\bibfield{author}{\bibinfo{person}{J{\"u}rgen Hackl} {and}
  \bibinfo{person}{Thibaut Dubernet}.} \bibinfo{year}{2019}\natexlab{}.
\newblock \showarticletitle{Epidemic spreading in urban areas using agent-based
  transportation models}.
\newblock \bibinfo{journal}{\emph{Future Internet}} \bibinfo{volume}{11},
  \bibinfo{number}{4} (\bibinfo{year}{2019}), \bibinfo{pages}{92}.
\newblock


\bibitem[\protect\citeauthoryear{Harris}{Harris}{2020}]%
        {harris2020subways}
\bibfield{author}{\bibinfo{person}{Jeffrey~E Harris}.}
  \bibinfo{year}{2020}\natexlab{}.
\newblock \showarticletitle{The subways seeded the massive coronavirus epidemic
  in new york city}.
\newblock \bibinfo{journal}{\emph{NBER Working Paper}} \bibinfo{number}{w27021}
  (\bibinfo{year}{2020}).
\newblock


\bibitem[\protect\citeauthoryear{Jaikishan~Patel}{Jaikishan~Patel}{2013}]%
        {ref:mumbai_rail_map}
\bibfield{author}{\bibinfo{person}{Mandar~Rane Jaikishan~Patel, Snehal~Patil}.}
  \bibinfo{year}{2013}\natexlab{}.
\newblock \bibinfo{title}{Mumbai Rail Map}.
\newblock
\newblock
\urldef\tempurl%
\url{https://commons.wikimedia.org/wiki/File:Mumbai_Rail_Map_-_English.jpg}
\showURL{%
\tempurl}


\bibitem[\protect\citeauthoryear{Leng, Whie, Hilton, Kucharski, Pellis, Stage,
  Davies, Keeling, and Flasche}{Leng et~al\mbox{.}}{2020}]%
        {leng2020effectiveness}
\bibfield{author}{\bibinfo{person}{Trystan Leng}, \bibinfo{person}{Connor
  Whie}, \bibinfo{person}{Joe Hilton}, \bibinfo{person}{Adam~J Kucharski},
  \bibinfo{person}{Lorenzo~J Pellis}, \bibinfo{person}{Helena Stage},
  \bibinfo{person}{Nicholas~G Davies}, \bibinfo{person}{Matt~J Keeling}, {and}
  \bibinfo{person}{Stefan Flasche}.} \bibinfo{year}{2020}\natexlab{}.
\newblock \showarticletitle{The effectiveness of social bubbles as part of a
  Covid-19 lockdown exit strategy, a modelling study}.
\newblock \bibinfo{journal}{\emph{medRxiv}} (\bibinfo{year}{2020}).
\newblock


\bibitem[\protect\citeauthoryear{Marathe and Vullikanti}{Marathe and
  Vullikanti}{2013}]%
        {marathe2013computational}
\bibfield{author}{\bibinfo{person}{Madhav Marathe} {and} \bibinfo{person}{Anil
  Kumar~S Vullikanti}.} \bibinfo{year}{2013}\natexlab{}.
\newblock \showarticletitle{Computational epidemiology}.
\newblock \bibinfo{journal}{\emph{Commun. ACM}} \bibinfo{volume}{56},
  \bibinfo{number}{7} (\bibinfo{year}{2013}), \bibinfo{pages}{88--96}.
\newblock


\bibitem[\protect\citeauthoryear{Mei, Chen, Zhu, Lees, Boukhanovsky, and
  Sloot}{Mei et~al\mbox{.}}{2015}]%
        {mei2015simulating}
\bibfield{author}{\bibinfo{person}{Shan Mei}, \bibinfo{person}{Bin Chen},
  \bibinfo{person}{Yifan Zhu}, \bibinfo{person}{Michael~Harold Lees},
  \bibinfo{person}{AV Boukhanovsky}, {and} \bibinfo{person}{Peter~MA Sloot}.}
  \bibinfo{year}{2015}\natexlab{}.
\newblock \showarticletitle{Simulating city-level airborne infectious
  diseases}.
\newblock \bibinfo{journal}{\emph{Computers, Environment and Urban Systems}}
  \bibinfo{volume}{51} (\bibinfo{year}{2015}), \bibinfo{pages}{97--105}.
\newblock


\bibitem[\protect\citeauthoryear{Nag}{Nag}{2017}]%
        {ref:mumbai_avg_speed}
\bibfield{author}{\bibinfo{person}{Shantonil Nag}.}
  \bibinfo{year}{2017}\natexlab{}.
\newblock \bibinfo{title}{Traffic speeds: India’s fastest \& slowest cities}.
\newblock
\newblock
\urldef\tempurl%
\url{https://www.cartoq.com/traffic-speeds-indias-fastest-slowest-cities/}
\showURL{%
\tempurl}


\bibitem[\protect\citeauthoryear{Nielsen and Sneppen}{Nielsen and
  Sneppen}{2020}]%
        {nielsen2020covid}
\bibfield{author}{\bibinfo{person}{Bjarke~Frost Nielsen} {and}
  \bibinfo{person}{Kim Sneppen}.} \bibinfo{year}{2020}\natexlab{}.
\newblock \showarticletitle{COVID-19 superspreading suggests mitigation by
  social network modulation}.
\newblock \bibinfo{journal}{\emph{medRxiv}} (\bibinfo{year}{2020}).
\newblock


\bibitem[\protect\citeauthoryear{Organization}{Organization}{2020}]%
        {whoPressRelease}
\bibfield{author}{\bibinfo{person}{Wold~Health Organization}.}
  \bibinfo{year}{2020}\natexlab{}.
\newblock \bibinfo{title}{WHO Director-General's opening remarks at the media
  briefing on COVID-19}.
\newblock
\newblock
\urldef\tempurl%
\url{https://www.who.int/dg/speeches/detail/who-director-general-s-opening-remarks-at-the-media-briefing-on-covid-19---16-march-2020}
\showURL{%
\tempurl}


\bibitem[\protect\citeauthoryear{Pastor-Satorras, Castellano, Van~Mieghem, and
  Vespignani}{Pastor-Satorras et~al\mbox{.}}{2015}]%
        {pastor2015epidemic}
\bibfield{author}{\bibinfo{person}{Romualdo Pastor-Satorras},
  \bibinfo{person}{Claudio Castellano}, \bibinfo{person}{Piet Van~Mieghem},
  {and} \bibinfo{person}{Alessandro Vespignani}.}
  \bibinfo{year}{2015}\natexlab{}.
\newblock \showarticletitle{Epidemic processes in complex networks}.
\newblock \bibinfo{journal}{\emph{Reviews of modern physics}}
  \bibinfo{volume}{87}, \bibinfo{number}{3} (\bibinfo{year}{2015}),
  \bibinfo{pages}{925}.
\newblock


\bibitem[\protect\citeauthoryear{Pinto, Magalhaes, Figueiredo, Alves, and
  Segura-Angel}{Pinto et~al\mbox{.}}{2020}]%
        {pinto2020local}
\bibfield{author}{\bibinfo{person}{Jose Paulo~Guedes Pinto},
  \bibinfo{person}{Patricia~Camargo Magalhaes}, \bibinfo{person}{Gerusa~Maria
  Figueiredo}, \bibinfo{person}{Domingos Alves}, {and}
  \bibinfo{person}{Diana~Maritza Segura-Angel}.}
  \bibinfo{year}{2020}\natexlab{}.
\newblock \showarticletitle{Local protection bubbles: an interpretation of the
  decrease in the velocity of coronavirus's spread in the city of Sao Paulo}.
\newblock \bibinfo{journal}{\emph{medRxiv}} (\bibinfo{year}{2020}).
\newblock


\bibitem[\protect\citeauthoryear{Sambaturu, Adhikari, Prakash, Venkatramanan,
  and Vullikanti}{Sambaturu et~al\mbox{.}}{2020}]%
        {sambaturu2020designing}
\bibfield{author}{\bibinfo{person}{Prathyush Sambaturu},
  \bibinfo{person}{Bijaya Adhikari}, \bibinfo{person}{B~Aditya Prakash},
  \bibinfo{person}{Srinivasan Venkatramanan}, {and} \bibinfo{person}{Anil
  Vullikanti}.} \bibinfo{year}{2020}\natexlab{}.
\newblock \showarticletitle{Designing Effective and Practical Interventions to
  Contain Epidemics}. In \bibinfo{booktitle}{\emph{Proceedings of the 19th
  International Conference on Autonomous Agents and MultiAgent Systems (AAMAS'
  20)}}. \bibinfo{publisher}{International Foundation for Autonomous Agents and
  Multiagent Systems}, \bibinfo{address}{Richland, SC},
  \bibinfo{pages}{1187--1195}.
\newblock


\bibitem[\protect\citeauthoryear{Singh, Marathe, Marathe, and Swarup}{Singh
  et~al\mbox{.}}{2018}]%
        {singh2018behavior}
\bibfield{author}{\bibinfo{person}{Meghendra Singh}, \bibinfo{person}{Achla
  Marathe}, \bibinfo{person}{Madhav~V Marathe}, {and} \bibinfo{person}{Samarth
  Swarup}.} \bibinfo{year}{2018}\natexlab{}.
\newblock \showarticletitle{Behavior model calibration for epidemic
  simulations}. In \bibinfo{booktitle}{\emph{Proceedings of the 17th
  International Conference on Autonomous Agents and MultiAgent Systems (AAMAS'
  18)}}. \bibinfo{publisher}{International Foundation for Autonomous Agents and
  Multiagent Systems}, \bibinfo{address}{Richland, SC},
  \bibinfo{pages}{1640--1648}.
\newblock


\bibitem[\protect\citeauthoryear{Swarup, Eubank, and Marathe}{Swarup
  et~al\mbox{.}}{2014}]%
        {swarup2014computational}
\bibfield{author}{\bibinfo{person}{Samarth Swarup}, \bibinfo{person}{Stephen~G
  Eubank}, {and} \bibinfo{person}{Madhav~V Marathe}.}
  \bibinfo{year}{2014}\natexlab{}.
\newblock \showarticletitle{Computational epidemiology as a challenge domain
  for multiagent systems}. In \bibinfo{booktitle}{\emph{Proceedings of the 2014
  international conference on Autonomous agents and MultiAgent systems (AAMAS'
  14)}}. \bibinfo{publisher}{International Foundation for Autonomous Agents and
  Multiagent Systems}, \bibinfo{address}{Richland, SC},
  \bibinfo{pages}{1173--1176}.
\newblock


\bibitem[\protect\citeauthoryear{Tizzani, Lenti, Ubaldi, Vezzani, Castellano,
  and Burioni}{Tizzani et~al\mbox{.}}{2018}]%
        {tizzani2018epidemic}
\bibfield{author}{\bibinfo{person}{Michele Tizzani}, \bibinfo{person}{Simone
  Lenti}, \bibinfo{person}{Enrico Ubaldi}, \bibinfo{person}{Alessandro
  Vezzani}, \bibinfo{person}{Claudio Castellano}, {and}
  \bibinfo{person}{Raffaella Burioni}.} \bibinfo{year}{2018}\natexlab{}.
\newblock \showarticletitle{Epidemic spreading and aging in temporal networks
  with memory}.
\newblock \bibinfo{journal}{\emph{Physical Review E}} \bibinfo{volume}{98},
  \bibinfo{number}{6} (\bibinfo{year}{2018}), \bibinfo{pages}{062315}.
\newblock


\bibitem[\protect\citeauthoryear{Venkatramanan, Lewis, Chen, Higdon,
  Vullikanti, and Marathe}{Venkatramanan et~al\mbox{.}}{2018}]%
        {venkatramanan2018using}
\bibfield{author}{\bibinfo{person}{Srinivasan Venkatramanan},
  \bibinfo{person}{Bryan Lewis}, \bibinfo{person}{Jiangzhuo Chen},
  \bibinfo{person}{Dave Higdon}, \bibinfo{person}{Anil Vullikanti}, {and}
  \bibinfo{person}{Madhav Marathe}.} \bibinfo{year}{2018}\natexlab{}.
\newblock \showarticletitle{Using data-driven agent-based models for
  forecasting emerging infectious diseases}.
\newblock \bibinfo{journal}{\emph{Epidemics}}  \bibinfo{volume}{22}
  (\bibinfo{year}{2018}), \bibinfo{pages}{43--49}.
\newblock


\bibitem[\protect\citeauthoryear{Verity, Okell, Dorigatti, Winskill, Whittaker,
  Imai, Cuomo-Dannenburg, Thompson, Walker, Fu, et~al\mbox{.}}{Verity
  et~al\mbox{.}}{2020}]%
        {verity2020estimates}
\bibfield{author}{\bibinfo{person}{Robert Verity}, \bibinfo{person}{Lucy~C
  Okell}, \bibinfo{person}{Ilaria Dorigatti}, \bibinfo{person}{Peter Winskill},
  \bibinfo{person}{Charles Whittaker}, \bibinfo{person}{Natsuko Imai},
  \bibinfo{person}{Gina Cuomo-Dannenburg}, \bibinfo{person}{Hayley Thompson},
  \bibinfo{person}{Patrick~GT Walker}, \bibinfo{person}{Han Fu},
  {et~al\mbox{.}}} \bibinfo{year}{2020}\natexlab{}.
\newblock \showarticletitle{Estimates of the severity of coronavirus disease
  2019: a model-based analysis}.
\newblock \bibinfo{journal}{\emph{The Lancet Infectious Diseases}}
  \bibinfo{volume}{20}, \bibinfo{number}{6} (\bibinfo{year}{2020}),
  \bibinfo{pages}{669 -- 677}.
\newblock


\bibitem[\protect\citeauthoryear{Vestergaard and G{\'e}nois}{Vestergaard and
  G{\'e}nois}{2015}]%
        {vestergaard2015temporal}
\bibfield{author}{\bibinfo{person}{Christian~L Vestergaard} {and}
  \bibinfo{person}{Mathieu G{\'e}nois}.} \bibinfo{year}{2015}\natexlab{}.
\newblock \showarticletitle{Temporal gillespie algorithm: Fast simulation of
  contagion processes on time-varying networks}.
\newblock \bibinfo{journal}{\emph{PLoS Comput Biol}} \bibinfo{volume}{11},
  \bibinfo{number}{10} (\bibinfo{year}{2015}), \bibinfo{pages}{e1004579}.
\newblock


\bibitem[\protect\citeauthoryear{Wang, He, Gao, Chow, Ozbay, and Iyer}{Wang
  et~al\mbox{.}}{2020}]%
        {wang2020nyu}
\bibfield{author}{\bibinfo{person}{Ding Wang}, \bibinfo{person}{Brian~Yueshuai
  He}, \bibinfo{person}{Jingqin Gao}, \bibinfo{person}{Joseph~YJ Chow},
  \bibinfo{person}{Kaan Ozbay}, {and} \bibinfo{person}{Shri Iyer}.}
  \bibinfo{year}{2020}\natexlab{}.
\newblock \bibinfo{title}{Impact of COVID-19 Behavioral Inertia on Reopening
  Strategies for New York City Transit}.
\newblock
\newblock
\showeprint[arxiv]{2006.13368}


\end{thebibliography}


\end{document}